\documentclass[useAMS,usenatbib]{mn2e}
\usepackage{url,times,graphicx,amsmath,amsfonts,amssymb,aas_macros,color,epsfig,varioref,subfigure,comment,natbib}

\usepackage[normalem]{ulem}

\newcommand{\App}[1]{Appendix~\ref{#1}}
\newcommand{\Tab}[1]{Table~\ref{#1}}
\newcommand{\Sec}[1]{Section~\ref{#1}}
\newcommand{\Eq}[1]{Eq.~(\ref{#1})}
\newcommand{\Fig}[1]{Fig.~\ref{#1}}
\newcommand{\hMpc}{{\ifmmode{h^{-1}{\rm Mpc}}\else{$h^{-1}$Mpc}\fi}}
\newcommand{\hGpc}{{\ifmmode{h^{-1}{\rm Mpc}}\else{$h^{-1}$Gpc}\fi}}
\newcommand{\hkpc}{{\ifmmode{h^{-1}{\rm kpc}}\else{$h^{-1}$kpc}\fi}}

\newcommand{\hMsun}{{\ifmmode{h^{-1}{\rm {M_{\odot}}}}\else{$h^{-1}{\rm{M_{\odot}}}$}\fi}}
\newcommand{\Msun}{{\ifmmode{{\rm {M_{\odot}}}}\else{${\rm{M_{\odot}}}$}\fi}}

\def\LCDM{\ensuremath{\Lambda}CDM}
\def\vtan{$v_{\rm tan}$}

\def\mlg{$M_{\rm LG}$}
\def\std{RAN}
\def\cs{CS}
\def\nlgcs{314}
\def\nlgstd{2028}
\def\mmw{$M_{\rm MW}$}
\def\mmto{$M_{\rm M31}$}
\def\mw{MW}
\def\mto{M31}

\newcommand{\tm}{{\ifmmode{\tau_{M}}\else{$\tau_{M}$}\fi}}
\newcommand{\tf}{{\ifmmode{\tau_{F}}\else{$\tau_{F}$}\fi}}
\newcommand{\tz}{{\ifmmode{\tau_{0}}\else{$\tau_{0}$}\fi}}
\newcommand{\tha}{{\ifmmode{\tau_{1/2}}\else{$\tau_{1/2}$}\fi}}
\newcommand{\tqa}{{\ifmmode{\tau_{3/4}}\else{$\tau_{3/4}$}\fi}}

\title[On the Mass Assembly History of the Local Group]
{On the Mass Assembly History of the Local Group}
\author[Carlesi Edoardo]
{
Edoardo Carlesi $^{1}$,
\thanks{E-mail: ecarlesi@aip.de}
Yehuda Hoffman $^{2}$,
Stefan Gottl\"ober $^{1}$,
Noam I. Libeskind $^{1,3}$,\and
Alexander Knebe $^{4,5,6}$,
Gustavo Yepes $^{4,5}$,
Sergey V. Pilipenko $^{7}$
\\
\\
$^{1}$Leibniz-Institut f\"ur Astrophysik Potsdam (AIP), An der Sternwarte 16, D-14482 Potsdam, Germany\\
$^{2}$Racah Institute of Physics, Givat Ram, 91040 Jerusalem, Israel\\
$^{3}$University of Lyon, UCB Lyon 1, CNRS/IN2P3, IPN Lyon, 69622 Villeurbanne, France\\
$^{4}$Departamento de F\'isica Te\'{o}rica, M\'{o}dulo 8, Facultad de Ciencias, Universidad Aut\'{o}noma de Madrid, 28049 Madrid, Spain\\
$^{5}$Centro de Investigaci\'{o}n Avanzada en F\'isica Fundamental (CIAFF), Facultad de Ciencias, Universidad Aut\'{o}noma de Madrid, 
28049 Madrid, Spain\\
$^{6}$International Centre for Radio Astronomy Research, University of Western Australia, 35 Stirling Highway, Crawley, 
Western Australia 6009, Australia\\
$^{7}$Astro Space centre of Lebedev Physical Institute of Russian Academy of Sciences, Profsojuznaja st. 84/32, 117997 Moscow, Russia\\
}

\begin{document}

\date{Submitted XXXX May XXXX}
\pagerange{\pageref{firstpage}--\pageref{lastpage}} \pubyear{2019}
\maketitle
\label{firstpage}
\setlength{\topmargin}{-1.5cm}


\begin{abstract}
In this work an ensemble of simulated Local Group analogues is used to constrain the properties of the mass
assembly history of the Milky Way (\mw) and Andromeda (\mto) galaxies.
These objects have been obtained using the constrained simulation technique, 
which ensures that simulated LGs live within a large scale environment akin to the observed one.
	Our results are compared against a standard $\Lambda$ Cold Dark Matter (\LCDM) series of
	simulations which use the same
cosmological parameters. This allows us to single out the effects of the constraints on the results.
We find that (a) the median constrained merging histories for \mto\ and \mw\ live above the 
	standard ones at the 1-$\sigma$ level, 
(b) the median formation time takes place $\approx 0.5$ Gyr earlier than unconstrained values, while the latest major merger happens on 
average $1.5$ Gyr earlier 
and (c) the probability for both LG haloes to have experienced their last major merger 
	in the first half of the history of the Universe 
is $\approx 50\%$ higher for the constrained pairs.
These results have been estimated to be significant at the $99\%$ confidence 
level by means of a Kolmogorov-Simirnov test.
These results suggest that the particular environment in which the Milky Way and Andromeda form plays a role in 
shaping their properties, and favours earlier formation and last major merger time values in agreement with other observational 
and theoretical considerations. 
\end{abstract}


\section{Introduction}\label{sec:intro}
The two main galaxies of the Local Group (LG), 
known as the Milky Way (\mw) and Andromeda (\mto), play a crucial role in shaping our understanding of galaxies in general. 
Due to their proximity, the large amount of high-quality data available for these two disk galaxies,
has given rise to so--called near field cosmology \citep{Bland-Hawtorn:2006, Bland-Hawtorn:2014},
an approach that aims at extracting cosmological information by observing the cosmic near-field. 
A number of studies have shown that by analysing the LG through a cosmological lens it is possible to understand, test and constrain 
the paradigm of galaxy and structure formation, the $\Lambda$ Cold Dark Matter (\LCDM) model 
\citep{deRossi:2009, Boylan-Kolchin:2010, Boylan-Kolchin:2012, Zavala:2012, Forero-Romero:2013, Garrison:2014, Tollerud:2014}
as well as non-standard, alternative theories \citep{Elahi:2015, Penzo:2016, Garaldi:2016, Carlesi:2017b}.

The location of a dark matter (DM) halo within a specific environment in the cosmic web has been suggested to play
an important role in shaping galaxies properties, such as its morphological type \citep{Dressler:1980,Nuza:2014,Metuki:2015}  and 
star formation rates, which for \mw-like galaxies are affected by higher-density environments \citep{Creasey:2015}.
Moreover, the filamentary nature of the LG surroundings affects its evolution through ram pressure stripping \citep{BenitezLlambay:2012} on the one hand and
feeding it with cold streams of gas \citep{Aragon-Calvo:2016} on the other, 
while is also linked to the anisotropic distribution of the \mw\ and \mto\ satellites \citep{Libeskind:2015a}.
\\
\\
\noindent
In the hierarchical picture of structure formation, where smaller DM haloes gradually merge into larger ones \citep{White:1978},
the properties of galaxies may strongly depend on their host and the way it has accreted its mass
\citep{Parry:2009, Stinson:2010, vanDokkum:2013, Rodriguez-Gomez:2017}. 
In particular, the formation of disks such as 
those observed in \mw\ and \mto, in a \LCDM\ cosmology, is crucially linked to the mass assembly history (MAH) of their DM haloes 
\citep{Toth:1992, Brook:2005, Naab:2006, Forero-Romero:2011, Sales:2012}, since recent collisions with massive galaxies
can inhibit their formation
\citep{Brook:2004, Stewart:2008, Scannapieco:2009, Hammer:2012, Scannapieco:2015}; an observation which is also 
confirmed by the properties of the white dwarf luminosity functions \citep{Kilic:2017}.
Moreover, the angular momentum and stellar mass of the \mw\ suggest that no significant merger 
\footnote{A major merger is defined in this paper as a merger where the mass ratio between the two halos is at least $1:10$.}
took place in the last 8 to 10 Gyr \citep{Hammer:2007, Ruchti:2015}.
More recently, \citet{Helmi:2018} and \citet{Iorio:2019}, using Gaia data, could identify a large galaxy that merged with the \mw's main progenitor
around $10$ Gyrs ago; this now fully coalesced galaxy, called Gaia-Enceladus, would be the latest massive object to have merged 
with the Galaxy. 
The most recent merging event experienced by the MW is most likely represented by the 
dwarf spheroidal galaxy Sagittarius (Sgr), which might be responsible
the vertical oscillations of the disk \citep{Laporte:2018}; the actual value of its progenitor's mass is actually unknown though it could be as massive as
$10^{11} M_{\sun}$, in which case it would classify as a major merger according to our definition.
Moreover, from the point of view of the MW dark matter halo, there is an ongoing interaction with the Large Magellanic Cloud (LMC), 
whose mass is estimated to be as large as $20\%$ of the MW \citep{Penarrubia:2016}.
However, being on its first crossing of the viral radius \citep{Besla:2007, Kallivayalil:2013} the LMC did not have enough time to perturb the disk of the MW \citep{Mastropietro:2005}, 
though a future LMC-MW merger is quite likely inevitable and will affect the properties of the bulge and the stellar halo significantly \citep{Cautun:2018}.\\
\\
In the case of \mto, where both the properties of its bulge \citep{Hammer:2010} and possibly also the 
anisotropic distribution of the satellites \citep{Fouquet:2012, Hammer:2013}, suggest that a major merger took place
around $\approx 7-8$ Gyr ago at the latest, while a number of studies based primarily on the age of stars in M31's bulge, 
claim that such an event took place as early as $10-11$ Gyr ago \citep{Brown:2008,Saglia:2010,Dalcanton:2012}.
It needs to be noted, however, that in contrast with the 
previous works \citet{Hammer:2018} found that \mto's thick disk might have originated from 
a merger of a large satellite which fully coalesced only around $1.8-3$ GYrs ago.
In general, however, there is a broad, theoretically and observationally motivated, consensus that the LG is characterized by a substantially quiet merging history 
during the last few gigayears.
\\
\\
\noindent
Hence, it is fundamental to keep in mind this considerations on the environment and MAH of the LG 
when practicing near field cosmology using $N$-body simulations.

A popular way of addressing this kind of studies is represented by the constrained simulation (\cs) technique, 
that aims at reproducing the 
$z=0$ Universe, in order to allow for a direct comparison of the simulation's output and observations
\citep{Gottloeber:2010, Sorce:2014, WangCS:2016, Sorce:2016}. 
Based on these approach, \citet{Carlesi:2016a} have developed the so called \emph{Local Group factory}, 
a numerical pipeline that allows to produce pairs of LG-like haloes in a large scale environment which closely mimics the observed one, 
within the framework of the Constrained Local UniversE Simulation (CLUES) project \footnote{http://www.clues-project.org}.
This tool has been used to simulate a series of LGs, which are used in this work to reconstruct the merger history 
of \mw\ and \mto, comparing them to those obtained for a control-group of similar objects identified in a standard random-phase \LCDM\ simulation (\std).
The focus is placed on \mw\ and \mto, the most prominent members of the LG, and on their MAH, using two metrics to do so: 
the last major merger time (\tm) and the moment when each halo acquired half its $z=0$ mass, defined as its formation time (\tf).
Our approach in this work is close to the one of \citet{Forero-Romero:2011}, however, the significantly larger sample of constrained LGs
allows us to put our conclusions on a firmer statistical footing.
\\
\\

\noindent
This work is structured as follows. \Sec{sec:methods} contains a description of the simulations and the methods used, discussing also
the definition of LG in cosmological simulations.
In \Sec{sec:results} we compute the median MAHs for \mw\ and \mto\ and derive
distribution functions for \tf\ and \tm, highlighting the differences between the \cs\ and \std\ and discussing their consequences.
In \Sec{sec:conclusions} we summarize our main results, discussing their implications and their relation to the current observational
and theoretical status.


\section{Methods}\label{sec:methods}

\begin{center}
\begin{figure*}
$\begin{array}{ccc}
\includegraphics[width=5.6cm]{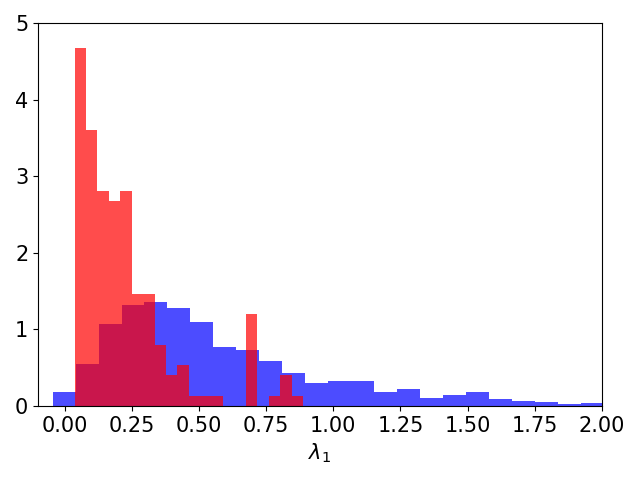} &
\includegraphics[width=5.6cm]{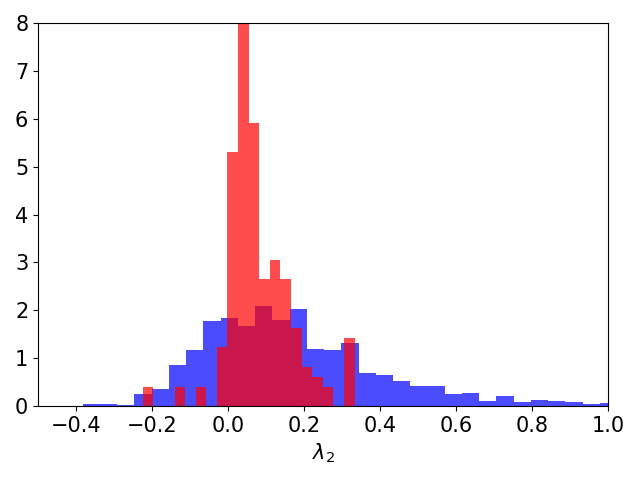} &
\includegraphics[width=5.6cm]{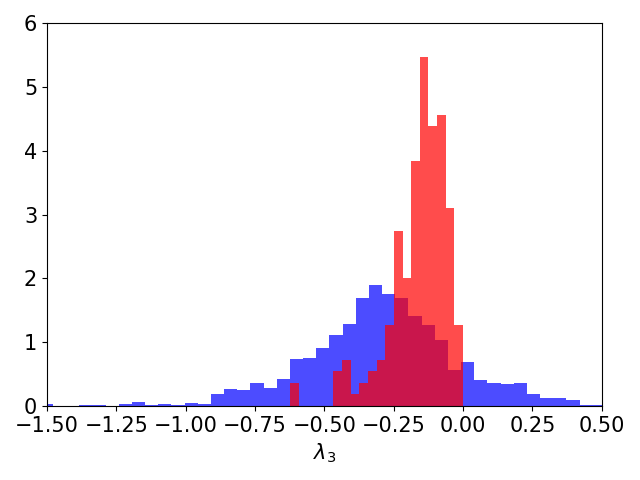} \\
\end{array}$
\caption{Distribution of the velocity shear tensor eigenvalues $\lambda_1, \lambda_2$ and $\lambda_3$ 
	computed at the position of the LG center of mass, for \cs\ (red color)
	and \std\ (in blue) samples. The area under the curve is normalized to 1.
	We note that the $\lambda$s are much more tightly distributed around the mean in \cs\ simulation than in 
	\std.
	\label{fig:evs}
	}
\end{figure*}
\end{center}

\begin{table}
\begin{center}
	\caption{Median and variance for the three shear tensor eigenvalues in \cs\ and \std.}
\label{tab:lambda}\begin{tabular}{ccc}
\hline
$\quad$ & \std & \cs \\
\hline
	$\lambda_1$ & $0.47\pm0.23$ & $0.16\pm0.03$ \\
	$\lambda_2$ & $0.14\pm0.23$ & $0.06\pm0.01$ \\
	$\lambda_3$ & $-0.29\pm0.07$ & $-0.14\pm0.01$ \\
\hline
\end{tabular}
\end{center}
\end{table}

This work is based on two different sets of $N$-body simulations, for standard \LCDM\ (labeled \std) and \cs s,
both run using Planck-I cosmological parameters: $\Omega _m = 0.31$, $\Omega _{\Lambda} = 0.69$, 
$h=0.67$ and $\sigma_8 = 0.83$ \citep{Planck:2013}, a box of $100$\hMpc, a mass resolution of $8 \times 10^7$ \hMsun\ and
a smoothing length of $2$\hkpc. For the \std\ simulation five $1024^3$ particle simulations where ran, while
due to the large number of \cs s needed to produce a significant sample size, 
a zoom-in technique was used for a $\approx7$\hMpc\ sphere around the center of a 100\hMpc\ box (with a resolution 
equivalent to $1024^3$ in the full box) using the \texttt{Ginnungagap} \footnote{https://github.com/ginnungagapgroup/ginnungagap} code. 
Each simulation has stored a total of $54$ snapshots, at constant separation of $250$ Myr from $z_{init} = 80$ to
$z=0$.
The following paragraphs will describe the main properties of the simulations and their relevance for the issue investigated here.

\subsection{The Simulations}

\textbf{Constrained simulations} 
These are cosmological simulations whose initial conditions (ICs) have been manipulated such that specific $z=0$ cosmographic constraints, set by observations of the local universe, are met. By ``cosmographic constraints'' we mean  local gravitational sources such as the Virgo cluster.
These constraints are given by the (grouped) Cosmic Flows 2 peculiar velocity data catalog \citep{Tully:2014},
using the procedure of \citet{Sorce:2015} to minimize observational biases.
Coupled with the Wiener filter technique and constrained realization of a Gaussian field algorithm \citep{Hoffman:1991}, these data
allow to reconstruct the velocity and density field at $z=0$. 
Constrained ICs are then generated by scaling the field back to the starting redshift ($z=80$ in our case) using the Reverse Zeldovich Approximation
described by \citet{Doumler:2013a, Doumler:2013b, Doumler:2013c}.
Peculiar velocity constraints are mostly effective on scales above $\approx 4$\hMpc , meaning that
below this threshold the random modes are expected to dominate. \citet{Carlesi:2016a} have shown that 
it is possible to optimize the above techniques to produce a numerical pipeline (the \emph{Local Group Factory}, LGF) 
that produces LG-like DM halo pairs at substantially
higher rate than expected by applying similar identification criteria to large cosmological volumes 
-- this despite the major role played by the random component on these scales.
This large sample of halo pairs, which lie in an environment akin to the observed one, can be used to 
constrain the dynamics of the LG, linking its mass and velocity to the cosmological and large scale context \citep{Carlesi:2016b, Carlesi:2017a}. 
In this work, we will use a subset of \nlgcs\ pairs selected from the first series of the LGF simulations 
according to criteria laid out in the following section.

The results drawn from these halo sample were compared to a second LG-like sample of objects, 
found in a standard random realization of the \LCDM\ model (\std). \\

\noindent
\textbf{Random simulations} In order to provide for a consistent control sample of dark matter haloes, a series of five 
full box simulations implementing exactly the same simulation setup
as the \cs s (apart from the application of zoom-in region).
In each of the simulations it was possible to identify $\sim 400$ LG-like halo pairs (using the criteria outlined in
\Sec{sec:lgm}) and resulting in a total of \nlgstd\ objects for the \std\ control sample.

\subsection{Halo Catalogs and Merger Trees}\label{sec:mah}
All the simulations have been analyzed using the same merger tree algorithm and halo finding software.
Halo catalogs are produced for each snapshot using the \texttt{AHF} spherical overdensity halo finder \citep{Knollmann:2009}.
The merger trees are derived using the \texttt{METROC++} code, 
which can be freely downloaded from \texttt{https://github.com/EdoardoCarlesi/MetroCPP}. 
\texttt{METROC++} stands for \emph{MErger TRees On C++}, and its main properties are described in \App{app:a}.
The merger trees obtained have been smoothed in the post-processing phase removing spurious mass accretion 
due to halo fly-bys.

\subsection{The Local Group Model}\label{sec:lgm}

The parameters and intervals used to define a simulated LG vary substantially among different authors
\citep[e.g. ][]{Forero-Romero:2013, Gonzalez:2014, Sawala:2014, Libeskind:2015a, Fattahi:2016a}, affecting at least to
some extent the conclusions reached.
It is therefore necessary to single out and motivate the specific choice made, which constitute a \emph{prior} or - using the
terminology introduced in \citet{Carlesi:2016b}, a Local Group model.
While any choice of the priors is generally related to the observational properties of the system, this needs to be
flexible enough to allow for a statistically significant sample of haloes to be built.
In this work, LG candidates are identified using the following criteria

\begin{itemize}
	\item Halo mass $> 0.5\times10^{12}$\hMsun 
	\item Total mass below $5.0\times 10^{12}$ \hMsun
	\item Halo separation in the range $(0.35 - 1.25)$ \hMpc
	\item Mass ratio \mmto / \mmw $< 3.0$
	\item Negative radial velocity
	\item Isolation, i.e., no third halo of mass $\geq$\mmw\ is located within $2.5$\hMpc\ from the center of mass of the LG
\end{itemize}

We focus only on the two largest LG members, 
labeling the least (most) massive member as \mw\ (\mto) \citep[see][]{Baiesi:2009, Karachentsev:2009, Diaz:2014}.
Given wildly different values for the tangential motion of the M31 in the literature \citep{Sohn:2012, Salomon:2016, Carlesi:2016b, 
Marel:2019}, we
allow for a large range of \vtan\ values, that are consistent with the state-of-the-art.
Moreover to avoid biasing our results by comparing different mass ranges, 
we ensure that the distributions of \mmw, \mmto\ and \mlg\ converge to the same median values in both samples as shown by 
\citep{Carlesi:2017a}.

\subsection{The environment}
The main difference between LG-like objects in the constrained simulation set is given by the \emph{environment}, which can be
classified using several different schemes such as those based on the tidal tensor \citep{Aragon-Calvo:2007, 
Hahn:2007, Forero-Romero:2009}, watershed segmentation \citep{Aragon-Calvo:2010}, 
the velocity shear tensor \citep{Hoffman:2012}, Bayesian reconstruction using tracers of the density field \citep{Leclercq:2015} 
among the others.
Here we implemented the classification algorithm of \citet{Hoffman:2012} which 
classifies the environment in terms of the eigenvalues $\lambda_i$ of the velocity shear tensor:

\begin{equation}\label{eq:shear}
\Sigma_{\alpha\beta} = - \frac{1}{2 H_0} \left ( \frac{\partial v_{\alpha}}{\partial r_{\beta}} - 
\frac{\partial v_{\beta}}{\partial r_{\alpha}} \right)
\end{equation}

\noindent
In this formalism, a given point in space is classified as a \emph{void} if the three eigenvalues are 
below a given threshold $\lambda_{th}$; it is labeled as a \emph{sheet} when two of eigenvalues are $<\lambda_{th}$, it is called
a \emph{filament} when one eigenvalue is $<\lambda_{th}$ and a \emph{knot} when all of the eigenvalues are $>\lambda_{th}$.
As shown in \citet{Carlesi:2016a} using a $2.5$\hMpc\ smoothing, these eigenvalues
are very well constrained at the position of the LG center of mass, where the environment is classified as a filament according
to our definition.
In this work, we used a $64^3$ regular grid and a $1.5$ \hMpc\ smoothing length in all the \std\ and \cs\ simulation
boxes to compute the $\lambda$s; in \Fig{fig:evs} we plot their values at the nearest grid point to each of the
simulated LG. 
This quantitatively shows the differences among the two samples, where the 
constrained nature of the \cs\ is reflected in the substantially smaller scatter.
The mean values together with their variance are shown in \Tab{tab:lambda}; these values are in very good agreement
with those obtained reconstructed by \citet{Libeskind:2015}.
Imposing the condition that the three eigenvalues around the LGs of the \std\ simulations 
simultaneously lie within $\pm 2\sigma$ around
the median \cs\ $\lambda_1, \lambda_2$ and $\lambda_3$ from \Tab{tab:lambda}, it turns out that (despite the large overlap
between the two distributions) only 8 out of 2028 pairs can be 
selected. Hence, while qualitatively \std\ and \cs\ samples live in similar kinds of environment, applying stricter quantitative 
criteria on the properties of the shear tensor around candidate local groups leads to a remarkable shrinking of the halo mple.


\begin{figure*}
\begin{center}
\includegraphics[width=14.5cm]{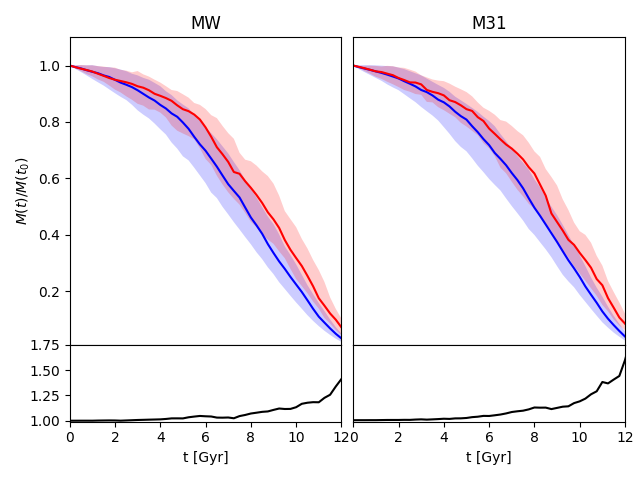} 
	\caption{Mass assembly histories for \mw\ and \mto\ for both the \cs\ (red) and \std\ (blue) samples; showing their 
	median values (thick lines) and their $25-{\rm th}$ and $75-{\rm th}$ percentiles (shaded region).
\label{fig:mmwm31_mah}}
\end{center}
\end{figure*}


\begin{center}
\begin{figure*}
\begin{tabular}{cc}
$\begin{array}{cc}
\vspace{-0.425cm}
\includegraphics[width=4.5cm, height=4.5cm]{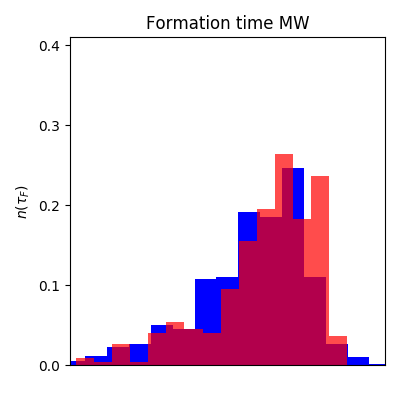} &
\hspace{-0.455cm}
\includegraphics[width=4.0cm, height=4.5cm]{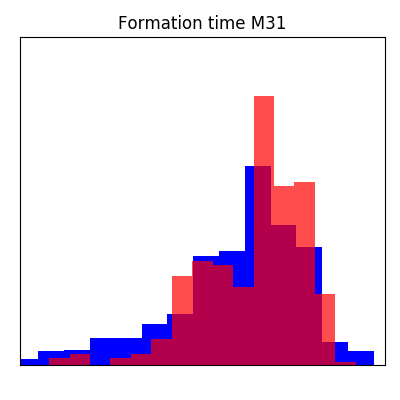} \\
\includegraphics[width=4.5cm, height=4.5cm]{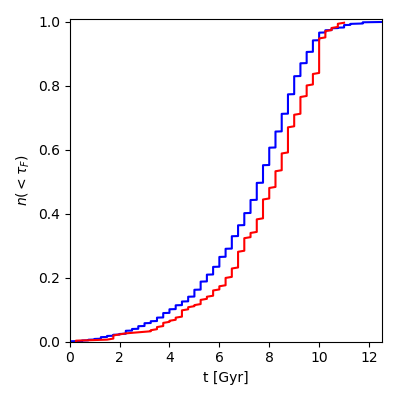} &
\hspace{-0.455cm}
\includegraphics[width=4.0cm, height=4.5cm]{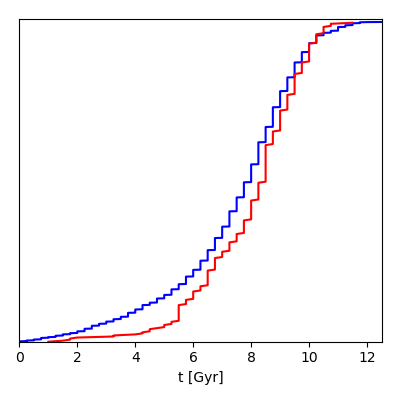}
\end{array}$ & 
$\begin{array}{cc}
\vspace{-0.425cm}
\includegraphics[width=4.5cm, height=4.5cm]{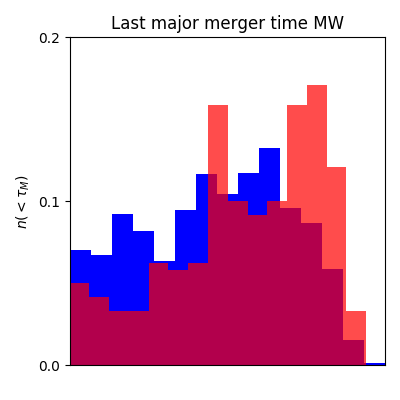} & 
\hspace{-0.455cm}
\includegraphics[width=4.0cm, height=4.5cm]{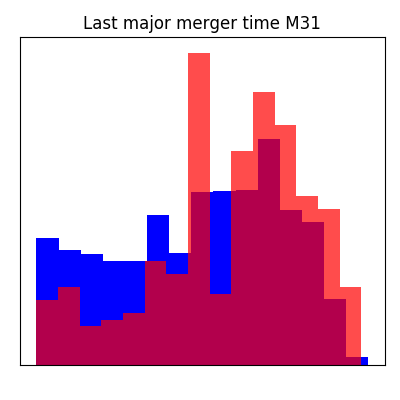} \\
\includegraphics[width=4.5cm, height=4.5cm]{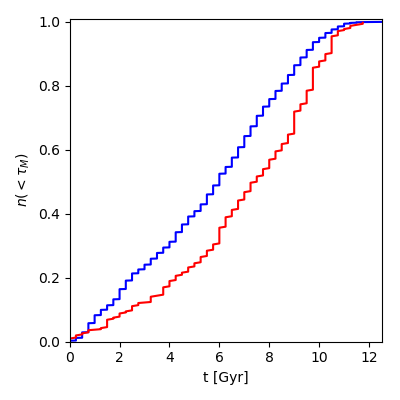} & 
\hspace{-0.455cm}
\includegraphics[width=4.0cm, height=4.5cm]{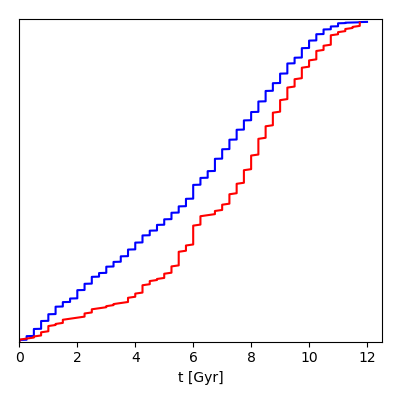} \\
\end{array}$
\end{tabular}
\caption{
	Formation and last major merger times (\tm) for \mw\ and \mto; histograms (upper panels) and cumulative distribution
	(lower panels). Red (blue) colors are for the \cs\ (\std) sample, while $x$-axis is in Gyr lookback time units. 
\label{fig:mmt_histo}}
\end{figure*}
\end{center}

\section{Results}\label{sec:results}
The sample of \nlgstd\ LG-like pairs identified in the \std\ simulations
functions as a benchmark to highlight the effect of the constrained environment on

\begin{itemize}
\item the mass assembly history
\item formation time
\item last major merger time
\end{itemize}

\noindent
that are going to be defined and discussed in the following subsections.

\subsection{Mass Assembly History}
\label{sec:mah1}

Our reconstruction of the halo history includes both smooth accretion of individual DM particles 
not bound into any discernible structure
other well resolved haloes. 
Using the \cs\ and \std\ halo samples and the LG model to identify \mw\ and \mto\ like-objects, 
we compute the \emph{median} MAH together with its $25$th and $75$th percentile intervals, shown in \Fig{fig:mmwm31_mah}.
We observe that, despite the significant overlap between the two distributions, in both \mw\ and \mto, 
the median values computed with the \cs\ sample differ significantly from the \std\ ones at the 1$\sigma$ level. 
Specifically, the median mass of LGs forming in constrained simulations, is larger at early times compared to ``random'' LGs.
This is a clear difference between the two samples, 
providing a first hint of the influence of the environment on the formation history of the 
LG. As can be seen in the last 6 Gyr, in fact, $\approx 50\%$ of the \cs\ LGs can be characterized by an MAH which lies above the 
$75$th percentile of the unconstrained \LCDM\ distribution. This means that haloes whose MAH lies within the residual $25$th percentile
of the distributions are expected to be seen twice as many times when introducing environmental constraints.
Such a finding is in qualitative agreement with the results of \citet{Forero-Romero:2011},
who also found that \cs\ MAHs consistently lied above the \std\ expectations. \\
\\

It is important to note that LGs identified in the \cs s, are not -- as a rule -- more massive than their \std\ counterparts. 
That they fulfill the same $z=0$ mass criteria is by construction. 
But it is important to note that at early times, at look back times of say $\sim10$ or 11 Gyr, 
the main progenitor of both the MW and M31 in both \cs\ and \std, have roughly the same mass of 0.15 
(in units of the $z=0 $ MW and M31 mass). 
In other words at the earliest epochs the samples have assembled similar amounts of their $z=0$ mass; 
the LGs in the CS grow faster at first and then their growth slows, 
while LGs in \std\ grow slower at first and then faster at late cosmic times, 
ensuring that by $z=0$ the samples fulfill the same mass criterium.

\begin{table}
\begin{center}
\caption{Best fit values for the $\alpha$ and $\beta$ parameters of \Eq{eq:mah} for \mw\ and \mto, obtained using the
\cs\ and \std\ LG samples.}
\label{tab:bestfit}\begin{tabular}{ccccc}
\hline
$\quad$ & \mw (\cs) & \mw (\std) & \mto (\cs) & \mto (\std) \\
\hline
	$\alpha$ & $5.54^{+0.24}_{-0.30}$ & $6.24^{+0.32}_{-0.45}$ & $6.01^{+0.22}_{-0.29}$ & $6.33^{+0.33}_{-0.27}$ \\  
	$\beta$ &  $2.71^{+0.18}_{-0.11}$ & $2.71^{+0.07}_{-0.23}$ & $2.66^{+0.16}_{-0.09}$ & $2.91^{+0.11}_{-0.14}$ \\
\hline
\end{tabular}
\end{center}
\end{table}

Many authors \citep[see e.g. ][]{Tasi:2004, McBride:2009, Boylan-Kolchin:2010} have found that the MAH of a DM halo 
is well fit by a modified exponential function of the kind 

\begin{equation}
\label{eq:mah}
\frac{M(z)}{M_0} = (1 +z)^{\beta} \exp(-\alpha(\sqrt{1+z} - 1))
\end{equation}

\noindent
We fit this function for \mw\ and M31 in the two halo samples, finding the parameters shown in \Tab{tab:bestfit}. 
In both cases, the difference between the numerical MAH and the analytical best-fit prediction is computed to be less than $2\%$.
As expected, the $\alpha$ values obtained for the \cs\ haloes are smaller than those found in \std\, 
reflecting the quieter accretion history that characterizes constrained \mw s and \mto s.
These results can be compared to the findings of \citet{Boylan-Kolchin:2010} and \citet{Forero-Romero:2011}.
These authors quoted a different set of best fit values, with $\alpha = 2.23$
and $\beta$, which they found to be $4.90$ and $4.50$, respectively. 
However, such a discrepancy can be expected since (a) the halo sample used by the aforementioned authors spans over a wider mass range, 
(b) the cosmological parameters were WMAP1 and WMAP5 instead of Planck-I (c) the FoF halo finding method is a major source
of differences in the MAH, with respect to the spherical overdensity \citep[see][]{Avila:2014} and 
(d) the function is fitted over a different redshift range (0 to 3 instead of 0 to 10).
We estimate the discrepancy between our values and the aforementioned ones computing root mean square difference ($\Delta_{RMS}$) of \Eq{eq:mah} 
over an interval $z=[0.0 - 6.0]$. 
Using different combinations of $\alpha$s and $\beta$s of \Tab{tab:bestfit} versus the best fit formulas of 
\citet{Boylan-Kolchin:2010} and \citet{Forero-Romero:2011} we see that $\Delta_{RMS}$ varies in the interval $0.08 - 0.28$ across the different combinations. 
These numbers are of the same order of magnitude of those obtained comparing the different MW/M31 (CS/RAN) $\alpha, \beta$ values of \Tab{tab:bestfit}, which give
e.g. $\Delta_{RMS} = 0.18$ in the case of MW(CS) and M31(RAN).
The implications of a flatter MAH for the expected values of \tm\ and \tf\ will be discussed in the following subsections.

\subsection{Formation Times}

\begin{table}
\begin{center}
\caption{Median values of the \tf\ (left table) and \tm\ (right table) distributions for \mw\ and \mto\ haloes, 
in the \cs\ and \std\ samples, with the relative $25$th and $75$th percentile intervals. Units are in Gyr lookback time
\label{tab:ft}}
\begin{tabular}{lll}

\begin{tabular}{c}
\multicolumn{1}{c}{$\quad$}\\
\hline
$\quad$ \\
\hline
\vspace{0.175cm}
\mw \\
\vspace{0.175cm}
\mto \\
\hline
\end{tabular} & 
\hspace{-0.75cm}

\begin{tabular}{cc}
\multicolumn{2}{c}{$\quad$\tf}\\
\hline
\cs  & \std  \\
\hline
\vspace{0.15cm}
$8.50_{-2.25}^{+0.75}$  & $8.00_{-1.75}^{+1.00}$ \\
\vspace{0.15cm}
$8.25_{-1.50}^{+1.00}$  & $7.75_{-1.75}^{+1.00}$ \\
\hline
\end{tabular} & 
\hspace{-0.75cm}

\begin{tabular}{cc}
\multicolumn{2}{c}{\tm}\\
\hline
\cs  & \std  \\
\hline
\vspace{0.15cm}
$7.75_{-2.25}^{+1.25}$ & $6.25_{-2.75}^{+2.00}$ \\
\vspace{0.15cm}
$7.50_{-2.25}^{+2.00}$ & $6.00_{-2.75}^{+2.00}$ \\
\hline
\end{tabular}

\end{tabular}
\end{center}
\end{table}

\begin{center}
\begin{table*}
	\caption{Fraction of haloes that formed (\tf) and experienced their latest major merger (\tm) 
	in the first half (\tha) and the first quarter (that is \tqa in lookback time) of the age of the Universe.
	The values are computed for individual \mw\ and \mto\ as well as for the whole LG, 
	i.e. when both haloes in the pair form or experience a merger at a time smaller than \tqa\ or \tha.
\label{tab:ft_tot}}
\begin{tabular}{ccccc}

\begin{tabular}{c}
\multicolumn{1}{c}{$\quad$} \\
\hline
$\quad$ \\
\hline
\vspace{0.15cm}
\mw \\
\vspace{0.15cm}
\mto \\
\vspace{0.15cm}
LG \\
\hline
\end{tabular} & 
\hspace{-0.75cm}

\begin{tabular}{cc}
\multicolumn{2}{c}{$n(\tf > \tha)$}\\
\hline
\cs & \std \\
\hline
\vspace{0.15cm}
$0.77$ & $0.67$ \\
\vspace{0.15cm}
$0.76$ & $0.71$ \\
\vspace{0.15cm}
$0.64$ & $0.49$ \\
\hline
\end{tabular} & 
\hspace{-0.75cm}

\begin{tabular}{cc}
\multicolumn{2}{c}{$n(\tf > \tqa)$}\\
\hline
\cs & \std \\
\hline
\vspace{0.15cm}
$0.16$ & $0.06$ \\
\vspace{0.15cm}
$0.12$ & $0.08$ \\
\vspace{0.15cm}
$0.06$ & $0.04$ \\
\hline
\end{tabular} &

\hspace{-0.75cm}
\begin{tabular}{cc}
	\multicolumn{2}{c}{$n(\tm > \tha)$}\\
\hline
\cs & \std \\
\hline
\vspace{0.15cm}
$0.60$ & $0.46$ \\
\vspace{0.15cm}
$0.58$ & $0.41$ \\
\vspace{0.15cm}
$0.32$ & $0.11$ \\
\hline
\end{tabular} & 

\hspace{-0.75cm}
\begin{tabular}{cc}
	\multicolumn{2}{c}{$n(\tm > \tqa)$}\\
\hline
\cs & \std \\
\hline
\vspace{0.15cm}
$0.18$ & $0.09$ \\
\vspace{0.15cm}
$0.17$ & $0.07$ \\
\vspace{0.15cm}
$0.07$ & $0.04$ \\
\hline
\end{tabular}

\end{tabular}
\end{table*}
\end{center}

	Halo formation time \tf\ is defined in this work as the time at which the main branch of the merger tree has reached half of the $z=0$ mass of the DM halo.
We compute this quantity for each of the LG in the \cs\ and \std\ samples and plot the \tf\ distribution in 
\Fig{fig:mmt_histo}. 
The median \tf\ values for all the distributions are shown in 
\Tab{tab:ft}.
In agreement with the findings of the previous subsection, \cs\ haloes are characterized by median \tf\ values which are 
$\approx 0.5$  Gyr above
the \std\ expectations. This can be seen in \Fig{fig:mmt_histo}, where the distributions are shifted at slightly higher lookback time values, 
with a more pronounced peak at $\approx8$ Gyr for both haloes. 
We complete the analysis computing the fraction of haloes that formed within the first half ($\tha = 6.9$ Gyr
lookback time) and within the first quarter ($\tqa = 10.35$ Gyr lookback time) of the age of the Universe.
In the first case, we find an excess probability $8\% - 15\%$ of \cs\ \mw s, \mto s and LGs 
to have formed in that time range; with a large overlap of of the values as shown in \Fig{fig:mmt_histo}.
In the second case, however, it can be noticed that the shift of the medians of the \cs\ distributions leaves a sizable amount of objects with 
a very early \tf, whose expected number is almost negligible the \std\ sample.
Nonetheless, the overall picture tells us that the environment-induced bias here is subdominant and the very large overlap between
the distributions coming from the different samples indicates that the \tf\ is essentially determined by the halo mass. 
In fact, it can be noticed that \mw s have earlier \tf s than \mto\ in both samples, consistent with the hierarchical formation 
picture and our LG model which prescribed \mmto $>$ \mmw.
\\
\\
\noindent
To conclude the section, we compare these results with the \tf\ distributions of \citet{Forero-Romero:2011}, 
who computed them using the Bolshoi simulation \citep{Klypin:2011}.
In that case, it turned out that individual haloes over a mass range comparable to the one of \mw\ and \mto\ in the \std\ simulation, 
for \tf\ lookback times $> 9$ Gyr, represent the $23\% - 29\%$ of the sample, and around $5\%$ of the total number of LG-like pairs. 
Applying this threshold to \std, we find that \mw\ and \mto\ have $\tf > 9$ Gyr in $26\%$ and $22\%$ of the times
respectively, while the share of LGs with both main haloes satisfying this criterion is $\approx 4\%$, showing that 
our benchmark rates broadly agree with those already existing in the literature.

\subsection{Last Major Merger}

Following \citet{Forero-Romero:2011}, 
we define as \emph{major merger} any merger event where mass of the accreted halo has a ratio of at least 1:10 with respect to the 
main one. 
The parameter \tm\ is defined as the age of the universe/lookback time when such an event last took place 
As for \tf, we show the distributions of both samples for \tm\ in \Fig{fig:mmt_histo}.
First, we notice that the median values presented in \Tab{tab:ft} show that \cs\ estimate are $1.5$ Gyr 
above the \std\ ones, i.e. $7.75$ for \mw\ and $7.50$ for \mto. 
This means that the expected merger history is consistently more quiet when simulating LGs in their correct environment.
The significance of this difference observed among the two distributions of the \tm\ values can be evaluated by means of a 
Kolmogorov-Smirnov (KS) test using the cumulative distributions shown in the lower panels of \Fig{fig:mmt_histo}.
The $p$-values for \mw\ and \mto\ are $2.42\times10^{-6}$ and $1.56\times10^{-4}$ respectively, so that the null hypothesis
that the two \tm\ samples come from the same distribution can be rejected at the $>99\%$ level. For comparison, the $p$-values
for the KS test in the case of \tf\ are both $\approx 0.21$.
We then estimate the effect of the sample size (which is substantially smaller for \cs) in the following way.
Randomly drawing 10000 subsamples of $N=\nlgcs$ from the \nlgstd\ \std\ pairs, we repeating the KS test for each one using 
the complete \std\ distribution, and look at the $p$-values obtained.
It is found that only less than $0.007\%$ of the times the \tm\ for the reduced \std\ subsample (for both \mw\ and \mto)
has a $p$-value smaller than $0.05$, allowing us to conclude that a random fluctuation is extremely unlikely to explain 
estimated the $p$-value.

To further quantify the incompatibility introduced by the constraints,
we focus again on the first half and the first quarter of the age of the Universe and compute the fractions of haloes whose \tm\ falls within the 
two intervals, which are shown in \Tab{tab:ft_tot}. 
Those haloes whose last major merger time falls within the first interval are said to have a \emph{quiet} merging history
whereas the second group has an \emph{extremely quiet} one. 
In the first case, the effect of the \cs\ is evident: the rates of \mw\ and \mto\ experiencing a last major merger in this time interval
is up to $20\%$ larger than \std\ rates. Moreover, more than a quarter of the LGs can be characterized by this kind of \tm s,
a value which is twice as large as
LGs identified in unconstrained \LCDM\ simulations, showing that the effect of the constraints is to 
force the sample towards a \emph{quiet} merger history.\\
\\
\noindent
On the other hand, the size of this effect is diminished in the case of haloes with an extremely quiet merger history. In this case the fractions 
of the samples are of comparable size between \std\ and \cs, with slightly higher values for the \cs\ haloes. Taking a perhaps more realistic approach, we can use two different \tm\ and compute probabilities for the combined \mw\ + \mto\ system for 
a more realistic set of priors. Analysing the angular momentum, stellar mass and properties of the disk of the \citet{Hammer:2007} concluded that 
\mw\ experienced its last major merger $\approx 10$ Gyr for \mw; while \citet{Hammer:2010} suggested that the bulge, disk and thick disk structure of 
lead to  $\tm \approx 8.75$ Gyr for \mto. Applying these two combined values it is found that such a configuration is realised in $\approx 6\%$
and $4\%$ of the times for the \cs\ and \std\ samples; the excess probability in this case is not as dramatic as the one found around the $8$ Gyr peak
and reveals that the environment does not lead to a thickening of the tail of the distribution corresponding to extremely quiet \tm\ values. 
The quietness of a MW MAH defined in this way is however at odds with the observations of the LMC, whose mass 
is most likely above $10\%$ of \mmw\ \citep{Penarrubia:2016} and is at its first crossing of the viral radius \citep{Besla:2007, Kallivayalil:2013}.
However, due to the accretion time being substantially smaller than the dynamical friction timescales, the interaction of the LMC with the Galaxy is 
still negligible and did not have an impact so far on its main structural properties \citep{Mastropietro:2005}.
In any case, our halo samples show that such a recent accretion, though being disfavoured, is also not ruled out either, 
and is expected to happen around $10\%$ in constrained simulations and $15\%$ in random ones.


\section{Conclusions}\label{sec:conclusions}

In this paper the mass assembly histories of the two defining members of the Local Group (LG) 
-- the \mw\ and \mto\ -- have been analyzed within the context of constrained simulations, 
that reproduce specific, observationally defined features of the large scale environment. 
Our method of constraining the initial conditions  in order to produce the $z=0$ environment 
is fairly accurate and is described in detail in \citet{Doumler:2013a, Sorce:2014, Carlesi:2016a}. 
In general, the most important cosmographic features on scales greater than $\sim$5 Mpc (such as the Virgo cluster, the Local Void, etc) 
are reproduced with the correct density and in the correct place. Moreover, the simulated LGs lie on a filament
whose properties (evaluated using the velocity shear tensor) are in good agreement with the reconstruction from the observational values
of \citet{Libeskind:2015}.
As smaller scales remain unconstrained, LGs are not automatically reproduced. Instead random haloes form embedded within the constrained environment. The criteria to define these as so-called ``LGs'' are termed a ``LG model''.

The first step is to introduce a LG model, namely a set of observationally motivated intervals for 
mass, mass ratio, isolation, relative velocity, and separation of two haloes at $z=0$. 
Any pair of haloes that meet the criteria defined in the LG model are -- for the purposes of this study --- considered a LG. The mass assembly histories (MAH) of the LGs that form in our constrained simulations are compared with halo pairs identified in un-constrained \LCDM\ simulations, run with the same Planck-I cosmological parameters, according to identical LG model.
The same halo finder as well as merger tree algorithms were used on both simulations to ensure consistency between the results.

We have identified a total of \nlgcs\ constrained LGs (\cs\ sample) and \nlgstd\ 
LG-like pairs in the standard \LCDM\ simulations (\std\ sample). 
We use these to characterize the MAH by computing formation times (\tf, 
defined as the lookback time when the main progenitor of the halo had acquired half its $z=0$ mass), 
last major merger times (\tm, defined as the time when the halo last had a merger of at least $1:10$) and
the growth curve of the \mw\ and \mto-like haloes.
This comparison can be used to isolate the effects of the environment on how LG like pairs of haloes acquire mass. Our main findings are 

\begin{itemize}
\item 
The median growth of a MW or M31 halo in our constrained simulations is faster at early times compared to haloes in unconstrained LGs. The difference is statistically significant at the 1$\sigma$ level. For example, MWs forming in constrained simulations acquire, in the median, 75\% of their $z=0$ mass at a lookback time of $\sim 7.25$ Gyr (similar values for M31). LGs identified in unconstrained simulations, acquire 75\% of their mass more than a Gyr later at look back times of $\sim$ 6Gyr.\\

\item As expected from above, an examination of the formation times (\tf), reveal a similar story:  their distribution for \cs\ LGs peak at values $\approx 0.5$ Gyr 
later than \std\ ones. We examine the probability that \emph{both} haloes have formed in the first half of the age of the Universe, 
which is $64\%$ in \cs\ compared to $49\%$ in \std. 
Formation times are only slightly affected by the environment, 
as the distribution functions of \tf\ overlap significantly between the \cs\ and \std\ samples. \\

\item The median last major merger times \tm\ are $7.75$ (\mw) and $7.5$ Gyr (\mto), $1.5$ Gyr above the
	\cs\ pairs. 
The difference \tm\ distributions obtained for \cs\ and \std\ 
		is significant at the $99\%$ confidence level, as given by the P-value computed with the KS-test.
\end{itemize}

\noindent
The above results are in agreement with observations \citep[e.g.][]{Hammer:2007, Ruchti:2015} which place this value within the $8-10$ Gyr interval for \mw\ and
close to the estimates of $\tm \approx 7-9$ Gyr \citep{Hammer:2010, Fouquet:2012, Hammer:2013} in the case of \mto.
Qualitatively, this tendency towards quieter merger histories associated with constrained simulations agrees with 
the findings of \citet{Forero-Romero:2011}, where a small sample of \cs s was used to derive an expected
$\tm> 10$ Gyr (lookback time) for both \mw\ and \mto. 
Extremely recent accretion of large satellites such as the LMC are found to take place around $10\%$ ($15\%$) of the times for \cs\ (\std) - a kind of event that would
classify as a major merger according to our definition, but would nonetheless leave the properties of the Galaxy largely unaffected.
\\

In order to examine our results in the context of empirically derived fitting formulas, we fit our MAH to \Eq{eq:mah}, taken from \cite{McBride:2009}. 
We find some differences with the best fit values of \citet{Boylan-Kolchin:2010} and \citet{Forero-Romero:2011}, which result in a $\Delta_{RMS}$ 
comprised within $0.08$ and $0.28$ depending on the best-fit parameter set used.
We believe these can be attributed to the different halo finding methods, cosmological parameters, LG models and redshift intervals used.

What we have attempted to accomplish in this work is to characterise the effect of cosmography on the history of the LG; 
to ask ``what role does environment play on producing a merger history for a pair of LG like haloes that is consistent with observations?''.
 For example, if 
morphological characteristics (such as the existence of  disks) of the \mw\ and \mto\ imply that no major merger took place
in \emph{either} of the main LG galaxies during at least the last $\approx 7$ Gyr, it can be said that we lived 
in a quiet merger history. 
LGs with quiet merger histories occur $\approx 27\%$ of the time in constrained environments but just 14\% of the time in unconstrained environments, albeit with the exact same set of criteria applied to identify a pair of haloes as a probability twice as much larger than the na\"ive \LCDM\ expectation, where 
pairs of DM haloes - defined \emph{via} the same LG model - experience this kind of merger history at a $14\%$ rate.
We may take an even more conservative approach, and consider only those LGs without any major mergers in the last $10.5$ Gyr. 
Roughly 6\% of constrained LGs and 4\% of unconstrained LGs satisfy such a criterium, a small - though not vanishing - probability and difference. 
To conclude, we have compared LG-like objects identified in constrained and in random simulations of the same cosmology 
using the same LG model, following the approach of \citet{Forero-Romero:2011} but using a larger statistical sample.
We have demonstrated that these samples are statistically different 
and therefore also the correct large scale environment as constructed in the constrained simulation 
plays an important role to understand the formation and evolution of our Local Group of galaxies.
\\
\\

\section*{Acknowledgements}

EC would like to thank Giorgio Mastrota for the support and the interesting discussions.
YH has been partially supported by the Israel Science Foundation (1358/18).
NIL acknowledges financial support of the Project IDEXLYON at the University of Lyon under the Investments
for the Future Program (ANR-16-IDEX-0005).
GY and AK are supported by Ministerio de Economia y Competitividad and the 
Fondo Europeo de Desarrollo Regional (MINECO/FEDER, UE) in Spain through grants 
AYA2015-63810-P and  PGC2018-094975-B-C21. AK is also supported by the Spanish Red Consolider MultiDark FPA2017-90566-REDC.
He further thanks Soundgarden for badmotorfinger.
SP is supported by the Russian Academy of Sciences program P-7 
RAS Program of basic research 12 "Problems of Origin and Evolution of the Universe".
The standard \std\ \LCDM\ simulations have been performed on the local cluster of the Leibniz Institut f\"ur Astrophysik in Potsdam.
We thank the Red Espa\~nola de Supercomputaci\' on   for  granting us 
computing time  in the Marenostrum Supercomputer at the BSC-CNS where the Local Group Factory simulations have been performed.


\bibliographystyle{mn2e}
\bibliography{biblio}

\begin{thebibliography}{}

\bibitem[\protect\citeauthoryear{{Aragon-Calvo}, {Neyrinck} \&
  {Silk}}{{Aragon-Calvo} et~al.}{2016}]{Aragon-Calvo:2016}
{Aragon-Calvo} M.~A.,  {Neyrinck} M.~C.,    {Silk} J.,  2016, arXiv e-prints,
  p. arXiv:1607.07881

\bibitem[\protect\citeauthoryear{{Arag{\'o}n-Calvo}, {Platen}, {van de
  Weygaert} \& {Szalay}}{{Arag{\'o}n-Calvo} et~al.}{2010}]{Aragon-Calvo:2010}
{Arag{\'o}n-Calvo} M.~A.,  {Platen} E.,  {van de Weygaert} R.,    {Szalay}
  A.~S.,  2010, \apj, 723, 364

\bibitem[\protect\citeauthoryear{{Arag{\'o}n-Calvo}, {van de Weygaert}, {Jones}
  \& {van der Hulst}}{{Arag{\'o}n-Calvo} et~al.}{2007}]{Aragon-Calvo:2007}
{Arag{\'o}n-Calvo} M.~A.,  {van de Weygaert} R.,  {Jones} B.~J.~T.,    {van der
  Hulst} J.~M.,  2007, \apjl, 655, L5

\bibitem[\protect\citeauthoryear{{Avila}, {Knebe}, {Pearce}, {Schneider},
  {Srisawat}, {Thomas}, {Behroozi}, {Elahi}, {Han}, {Mao}, {Onions},
  {Rodriguez-Gomez} \& {Tweed}}{{Avila} et~al.}{2014}]{Avila:2014}
{Avila} S.,  {Knebe} A.,  {Pearce} F.~R.,  {Schneider} A.,  {Srisawat} C.,
  {Thomas} P.~A.,  {Behroozi} P.,  {Elahi} P.~J.,  {Han} J.,  {Mao} Y.-Y.,
  {Onions} J.,  {Rodriguez-Gomez} V.,    {Tweed} D.,  2014, \mnras, 441, 3488

\bibitem[\protect\citeauthoryear{{Baiesi Pillastrini}}{{Baiesi
  Pillastrini}}{2009}]{Baiesi:2009}
{Baiesi Pillastrini} G.~C.,  2009, \mnras, 397, 1990

\bibitem[\protect\citeauthoryear{Benitez-Llambay, Navarro, Abadi, Gottloeber,
  Yepes, Hoffman \& Steinmetz}{Benitez-Llambay
  et~al.}{2013}]{BenitezLlambay:2012}
Benitez-Llambay A.,  Navarro J.~F.,  Abadi M.~G.,  Gottloeber S.,  Yepes G.,
  Hoffman Y.,    Steinmetz M.,  2013, Astrophys. J., 763, L41

\bibitem[\protect\citeauthoryear{{Besla}, {Kallivayalil}, {Hernquist},
  {Robertson}, {Cox}, {van der Marel} \& {Alcock}}{{Besla}
  et~al.}{2007}]{Besla:2007}
{Besla} G.,  {Kallivayalil} N.,  {Hernquist} L.,  {Robertson} B.,  {Cox} T.~J.,
   {van der Marel} R.~P.,    {Alcock} C.,  2007, \apj, 668, 949

\bibitem[\protect\citeauthoryear{{Bland-Hawthorn} \&
  {Freeman}}{{Bland-Hawthorn} \& {Freeman}}{2014}]{Bland-Hawtorn:2014}
{Bland-Hawthorn} J.,  {Freeman} K.,  2014, The Origin of the Galaxy and Local
  Group, Saas-Fee Advanced Course, Volume 37.~ISBN
  978-3-642-41719-1.~Springer-Verlag Berlin Heidelberg, 2014, p.~1, 37, 1

\bibitem[\protect\citeauthoryear{{Bland-Hawthorn} \&
  {Peebles}}{{Bland-Hawthorn} \& {Peebles}}{2006}]{Bland-Hawtorn:2006}
{Bland-Hawthorn} J.,  {Peebles} P.~J.~E.,  2006, Science, 313, 311

\bibitem[\protect\citeauthoryear{{Boylan-Kolchin}, {Bullock} \&
  {Kaplinghat}}{{Boylan-Kolchin} et~al.}{2012}]{Boylan-Kolchin:2012}
{Boylan-Kolchin} M.,  {Bullock} J.~S.,    {Kaplinghat} M.,  2012, \mnras, 422,
  1203

\bibitem[\protect\citeauthoryear{{Boylan-Kolchin}, {Springel}, {White} \&
  {Jenkins}}{{Boylan-Kolchin} et~al.}{2010}]{Boylan-Kolchin:2010}
{Boylan-Kolchin} M.,  {Springel} V.,  {White} S.~D.~M.,    {Jenkins} A.,  2010,
  \mnras, 406, 896

\bibitem[\protect\citeauthoryear{{Brook}, {Gibson}, {Martel} \&
  {Kawata}}{{Brook} et~al.}{2005}]{Brook:2005}
{Brook} C.~B.,  {Gibson} B.~K.,  {Martel} H.,    {Kawata} D.,  2005, \apj, 630,
  298

\bibitem[\protect\citeauthoryear{{Brook}, {Kawata}, {Gibson} \&
  {Freeman}}{{Brook} et~al.}{2004}]{Brook:2004}
{Brook} C.~B.,  {Kawata} D.,  {Gibson} B.~K.,    {Freeman} K.~C.,  2004, \apj,
  612, 894

\bibitem[\protect\citeauthoryear{{Brown} et~al.,}{{Brown}
  et~al.}{2008}]{Brown:2008}
{Brown} T.~M.,  et~al., 2008, \apjl, 685, L121

\bibitem[\protect\citeauthoryear{{Carlesi}, {Hoffman}, {Sorce} \&
  {Gottl{\"o}ber}}{{Carlesi} et~al.}{2017a}]{Carlesi:2017a}
{Carlesi} E.,  {Hoffman} Y.,  {Sorce} J.~G.,    {Gottl{\"o}ber} S.,  {2017a},
  \mnras, 465, 4886

\bibitem[\protect\citeauthoryear{{Carlesi}, {Hoffman}, {Sorce}, {Gottloeber},
  {Yepes}, {Courtois} \& {Tully}}{{Carlesi} et~al.}{2016b}]{Carlesi:2016b}
{Carlesi} E.,  {Hoffman} Y.,  {Sorce} J.~G.,  {Gottloeber} S.,  {Yepes} G.,
  {Courtois} H.,    {Tully} R.~B.,  2016b, \mnras, 460, L5

\bibitem[\protect\citeauthoryear{{Carlesi}, {Mota} \& {Winther}}{{Carlesi}
  et~al.}{2017b}]{Carlesi:2017b}
{Carlesi} E.,  {Mota} D.~F.,    {Winther} H.~A.,  {2017b}, \mnras, 466, 4813

\bibitem[\protect\citeauthoryear{{Carlesi}, {Sorce}, {Hoffman},
  {Gottl{\"o}ber}, {Yepes}, {Libeskind}, {Pilipenko}, {Knebe}, {Courtois},
  {Tully} \& {Steinmetz}}{{Carlesi} et~al.}{2016a}]{Carlesi:2016a}
{Carlesi} E.,  {Sorce} J.~G.,  {Hoffman} Y.,  {Gottl{\"o}ber} S.,  {Yepes} G.,
  {Libeskind} N.~I.,  {Pilipenko} S.~V.,  {Knebe} A.,  {Courtois} H.,  {Tully}
  R.~B.,    {Steinmetz} M.,  2016a, \mnras, 458, 900

\bibitem[\protect\citeauthoryear{Cautun, Deason, Frenk \& McAlpine}{Cautun
  et~al.}{2018}]{Cautun:2018}
Cautun M.,  Deason A.~J.,  Frenk C.~S.,    McAlpine S.,  2018, Monthly Notices
  of the Royal Astronomical Society, 483, 2185

\bibitem[\protect\citeauthoryear{{Creasey}, {Scannapieco}, {Nuza}, {Yepes},
  {Gottl{\"o}ber} \& {Steinmetz}}{{Creasey} et~al.}{2015}]{Creasey:2015}
{Creasey} P.,  {Scannapieco} C.,  {Nuza} S.~E.,  {Yepes} G.,  {Gottl{\"o}ber}
  S.,    {Steinmetz} M.,  2015, \apjl, 800, L4

\bibitem[\protect\citeauthoryear{{Dalcanton} et~al.,}{{Dalcanton}
  et~al.}{2012}]{Dalcanton:2012}
{Dalcanton} J.~J.,  et~al., 2012, \apjs, 200, 18

\bibitem[\protect\citeauthoryear{{de Rossi}, {Tissera}, {De Lucia} \&
  {Kauffmann}}{{de Rossi} et~al.}{2009}]{deRossi:2009}
{de Rossi} M.~E.,  {Tissera} P.~B.,  {De Lucia} G.,    {Kauffmann} G.,  2009,
  \mnras, 395, 210

\bibitem[\protect\citeauthoryear{{Diaz}, {Koposov}, {Irwin}, {Belokurov} \&
  {Evans}}{{Diaz} et~al.}{2014}]{Diaz:2014}
{Diaz} J.~D.,  {Koposov} S.~E.,  {Irwin} M.,  {Belokurov} V.,    {Evans} N.~W.,
   2014, \mnras, 443, 1688

\bibitem[\protect\citeauthoryear{{Doumler}, {Courtois}, {Gottl{\"o}ber} \&
  {Hoffman}}{{Doumler} et~al.}{2013b}]{Doumler:2013b}
{Doumler} T.,  {Courtois} H.,  {Gottl{\"o}ber} S.,    {Hoffman} Y.,  {2013b},
  \mnras, 430, 902

\bibitem[\protect\citeauthoryear{{Doumler}, {Gottl{\"o}ber}, {Hoffman} \&
  {Courtois}}{{Doumler} et~al.}{2013c}]{Doumler:2013c}
{Doumler} T.,  {Gottl{\"o}ber} S.,  {Hoffman} Y.,    {Courtois} H.,  {2013c},
  \mnras, 430, 912

\bibitem[\protect\citeauthoryear{{Doumler}, {Hoffman}, {Courtois} \&
  {Gottl{\"o}ber}}{{Doumler} et~al.}{2013a}]{Doumler:2013a}
{Doumler} T.,  {Hoffman} Y.,  {Courtois} H.,    {Gottl{\"o}ber} S.,  {2013a},
  \mnras, 430, 888

\bibitem[\protect\citeauthoryear{{Dressler}}{{Dressler}}{1980}]{Dressler:1980}
{Dressler} A.,  1980, \apj, 236, 351

\bibitem[\protect\citeauthoryear{{Elahi}, {Lewis}, {Power}, {Carlesi} \&
  {Knebe}}{{Elahi} et~al.}{2015}]{Elahi:2015}
{Elahi} P.~J.,  {Lewis} G.~F.,  {Power} C.,  {Carlesi} E.,    {Knebe} A.,
  2015, \mnras, 452, 1341

\bibitem[\protect\citeauthoryear{{Fattahi}, {Navarro}, {Sawala}, {Frenk},
  {Oman}, {Crain}, {Furlong}, {Schaller}, {Schaye}, {Theuns} \&
  {Jenkins}}{{Fattahi} et~al.}{2016}]{Fattahi:2016a}
{Fattahi} A.,  {Navarro} J.~F.,  {Sawala} T.,  {Frenk} C.~S.,  {Oman} K.~A.,
  {Crain} R.~A.,  {Furlong} M.,  {Schaller} M.,  {Schaye} J.,  {Theuns} T.,
  {Jenkins} A.,  2016, \mnras, 457, 844

\bibitem[\protect\citeauthoryear{{Forero-Romero}, {Hoffman}, {Bustamante},
  {Gottl{\"o}ber} \& {Yepes}}{{Forero-Romero}
  et~al.}{2013}]{Forero-Romero:2013}
{Forero-Romero} J.~E.,  {Hoffman} Y.,  {Bustamante} S.,  {Gottl{\"o}ber} S.,
  {Yepes} G.,  2013, \apjl, 767, L5

\bibitem[\protect\citeauthoryear{{Forero-Romero}, {Hoffman}, {Gottl{\"o}ber},
  {Klypin} \& {Yepes}}{{Forero-Romero} et~al.}{2009}]{Forero-Romero:2009}
{Forero-Romero} J.~E.,  {Hoffman} Y.,  {Gottl{\"o}ber} S.,  {Klypin} A.,
  {Yepes} G.,  2009, \mnras, 396, 1815

\bibitem[\protect\citeauthoryear{{Forero-Romero}, {Hoffman}, {Yepes},
  {Gottl{\"o}ber}, {Piontek}, {Klypin} \& {Steinmetz}}{{Forero-Romero}
  et~al.}{2011}]{Forero-Romero:2011}
{Forero-Romero} J.~E.,  {Hoffman} Y.,  {Yepes} G.,  {Gottl{\"o}ber} S.,
  {Piontek} R.,  {Klypin} A.,    {Steinmetz} M.,  2011, \mnras, 417, 1434

\bibitem[\protect\citeauthoryear{{Fouquet}, {Hammer}, {Yang}, {Puech} \&
  {Flores}}{{Fouquet} et~al.}{2012}]{Fouquet:2012}
{Fouquet} S.,  {Hammer} F.,  {Yang} Y.,  {Puech} M.,    {Flores} H.,  2012,
  \mnras, 427, 1769

\bibitem[\protect\citeauthoryear{{Garaldi}, {Baldi} \& {Moscardini}}{{Garaldi}
  et~al.}{2016}]{Garaldi:2016}
{Garaldi} E.,  {Baldi} M.,    {Moscardini} L.,  2016, \jcap, 1, 050

\bibitem[\protect\citeauthoryear{{Garrison-Kimmel}, {Boylan-Kolchin}, {Bullock}
  \& {Kirby}}{{Garrison-Kimmel} et~al.}{2014}]{Garrison:2014}
{Garrison-Kimmel} S.,  {Boylan-Kolchin} M.,  {Bullock} J.~S.,    {Kirby} E.~N.,
   2014, \mnras, 444, 222

\bibitem[\protect\citeauthoryear{{Gonz{\'a}lez}, {Kravtsov} \&
  {Gnedin}}{{Gonz{\'a}lez} et~al.}{2014}]{Gonzalez:2014}
{Gonz{\'a}lez} R.~E.,  {Kravtsov} A.~V.,    {Gnedin} N.~Y.,  2014, \apj, 793,
  91

\bibitem[\protect\citeauthoryear{Gottl{\"o}ber, Hoffman \& Yepes}{Gottl{\"o}ber
  et~al.}{2010}]{Gottloeber:2010}
Gottl{\"o}ber S.,  Hoffman Y.,    Yepes G.,  2010, {High Performance Computing
  in Science and Engineering}.
Springer Berlin Heidelberg, Berlin, Heidelberg, p. 309–322

\bibitem[\protect\citeauthoryear{{Hahn}, {Carollo}, {Porciani} \&
  {Dekel}}{{Hahn} et~al.}{2007}]{Hahn:2007}
{Hahn} O.,  {Carollo} C.~M.,  {Porciani} C.,    {Dekel} A.,  2007, \mnras, 381,
  41

\bibitem[\protect\citeauthoryear{{Hammer}, {Puech}, {Chemin}, {Flores} \&
  {Lehnert}}{{Hammer} et~al.}{2007}]{Hammer:2007}
{Hammer} F.,  {Puech} M.,  {Chemin} L.,  {Flores} H.,    {Lehnert} M.~D.,
  2007, \apj, 662, 322

\bibitem[\protect\citeauthoryear{{Hammer}, {Yang}, {Flores} \&
  {Puech}}{{Hammer} et~al.}{2012}]{Hammer:2012}
{Hammer} F.,  {Yang} Y.,  {Flores} H.,    {Puech} M.,  2012, Modern Physics
  Letters A, 27, 1230034

\bibitem[\protect\citeauthoryear{{Hammer}, {Yang}, {Fouquet}, {Pawlowski},
  {Kroupa}, {Puech}, {Flores} \& {Wang}}{{Hammer} et~al.}{2013}]{Hammer:2013}
{Hammer} F.,  {Yang} Y.,  {Fouquet} S.,  {Pawlowski} M.~S.,  {Kroupa} P.,
  {Puech} M.,  {Flores} H.,    {Wang} J.,  2013, \mnras, 431, 3543

\bibitem[\protect\citeauthoryear{{Hammer}, {Yang}, {Wang}, {Ibata}, {Flores} \&
  {Puech}}{{Hammer} et~al.}{2018}]{Hammer:2018}
{Hammer} F.,  {Yang} Y.~B.,  {Wang} J.~L.,  {Ibata} R.,  {Flores} H.,
  {Puech} M.,  2018, \mnras, 475, 2754

\bibitem[\protect\citeauthoryear{{Hammer}, {Yang}, {Wang}, {Puech}, {Flores} \&
  {Fouquet}}{{Hammer} et~al.}{2010}]{Hammer:2010}
{Hammer} F.,  {Yang} Y.~B.,  {Wang} J.~L.,  {Puech} M.,  {Flores} H.,
  {Fouquet} S.,  2010, \apj, 725, 542

\bibitem[\protect\citeauthoryear{{Helmi}, {Babusiaux}, {Koppelman}, {Massari},
  {Veljanoski} \& {Brown}}{{Helmi} et~al.}{2018}]{Helmi:2018}
{Helmi} A.,  {Babusiaux} C.,  {Koppelman} H.~H.,  {Massari} D.,  {Veljanoski}
  J.,    {Brown} A.~G.~A.,  2018, \nat, 563, 85

\bibitem[\protect\citeauthoryear{{Hoffman}, {Metuki}, {Yepes}, {Gottl{\"o}ber},
  {Forero-Romero}, {Libeskind} \& {Knebe}}{{Hoffman}
  et~al.}{2012}]{Hoffman:2012}
{Hoffman} Y.,  {Metuki} O.,  {Yepes} G.,  {Gottl{\"o}ber} S.,  {Forero-Romero}
  J.~E.,  {Libeskind} N.~I.,    {Knebe} A.,  2012, \mnras, 425, 2049

\bibitem[\protect\citeauthoryear{{Hoffman} \& {Ribak}}{{Hoffman} \&
  {Ribak}}{1991}]{Hoffman:1991}
{Hoffman} Y.,  {Ribak} E.,  1991, \apjl, 380, L5

\bibitem[\protect\citeauthoryear{{Iorio} \& {Belokurov}}{{Iorio} \&
  {Belokurov}}{2019}]{Iorio:2019}
{Iorio} G.,  {Belokurov} V.,  2019, \mnras, 482, 3868

\bibitem[\protect\citeauthoryear{{Kallivayalil}, {van der Marel}, {Besla},
  {Anderson} \& {Alcock}}{{Kallivayalil} et~al.}{2013}]{Kallivayalil:2013}
{Kallivayalil} N.,  {van der Marel} R.~P.,  {Besla} G.,  {Anderson} J.,
  {Alcock} C.,  2013, \apj, 764, 161

\bibitem[\protect\citeauthoryear{{Karachentsev}, {Kashibadze}, {Makarov} \&
  {Tully}}{{Karachentsev} et~al.}{2009}]{Karachentsev:2009}
{Karachentsev} I.~D.,  {Kashibadze} O.~G.,  {Makarov} D.~I.,    {Tully} R.~B.,
  2009, \mnras, 393, 1265

\bibitem[\protect\citeauthoryear{{Kilic}, {Munn}, {Harris}, {von Hippel},
  {Liebert}, {Williams}, {Jeffery} \& {DeGennaro}}{{Kilic}
  et~al.}{2017}]{Kilic:2017}
{Kilic} M.,  {Munn} J.~A.,  {Harris} H.~C.,  {von Hippel} T.,  {Liebert} J.~W.,
   {Williams} K.~A.,  {Jeffery} E.,    {DeGennaro} S.,  2017, \apj, 837, 162

\bibitem[\protect\citeauthoryear{{Klypin}, {Trujillo-Gomez} \&
  {Primack}}{{Klypin} et~al.}{2011}]{Klypin:2011}
{Klypin} A.~A.,  {Trujillo-Gomez} S.,    {Primack} J.,  2011, \apj, 740, 102

\bibitem[\protect\citeauthoryear{{Knebe} \& et al.}{{Knebe} \&
  et~al.}{2013}]{Knebe:2013}
{Knebe} A.,  et al. 2013, \mnras, 435, 1618

\bibitem[\protect\citeauthoryear{{Knollmann} \& {Knebe}}{{Knollmann} \&
  {Knebe}}{2009}]{Knollmann:2009}
{Knollmann} S.~R.,  {Knebe} A.,  2009, \apjs, 182, 608

\bibitem[\protect\citeauthoryear{{Laporte} \& et al.}{{Laporte} \&
  et~al.}{2018}]{Laporte:2018}
{Laporte} C. F.~P.,  et al. 2018, \mnras, 481, 286

\bibitem[\protect\citeauthoryear{{Leclercq}, {Jasche} \& {Wandelt}}{{Leclercq}
  et~al.}{2015}]{Leclercq:2015}
{Leclercq} F.,  {Jasche} J.,    {Wandelt} B.,  2015, \aap, 576, L17

\bibitem[\protect\citeauthoryear{{Libeskind}, {Hoffman}, {Tully}, {Courtois},
  {Pomar{\`e}de}, {Gottl{\"o}ber} \& {Steinmetz}}{{Libeskind}
  et~al.}{2015a}]{Libeskind:2015a}
{Libeskind} N.~I.,  {Hoffman} Y.,  {Tully} R.~B.,  {Courtois} H.~M.,
  {Pomar{\`e}de} D.,  {Gottl{\"o}ber} S.,    {Steinmetz} M.,  2015a, \mnras,
  452, 1052

\bibitem[\protect\citeauthoryear{{Libeskind}, {Hoffman}, {Tully}, {Courtois},
  {Pomar{\`e}de}, {Gottl{\"o}ber} \& {Steinmetz}}{{Libeskind}
  et~al.}{2015b}]{Libeskind:2015}
{Libeskind} N.~I.,  {Hoffman} Y.,  {Tully} R.~B.,  {Courtois} H.~M.,
  {Pomar{\`e}de} D.,  {Gottl{\"o}ber} S.,    {Steinmetz} M.,  2015b, \mnras,
  452, 1052

\bibitem[\protect\citeauthoryear{Mastropietro, Moore, Mayer, Wadsley \&
  Stadel}{Mastropietro et~al.}{2005}]{Mastropietro:2005}
Mastropietro C.,  Moore B.,  Mayer L.,  Wadsley J.,    Stadel J.,  2005,
  Monthly Notices of the Royal Astronomical Society, 363, 509

\bibitem[\protect\citeauthoryear{{McBride}, {Fakhouri} \& {Ma}}{{McBride}
  et~al.}{2009}]{McBride:2009}
{McBride} J.,  {Fakhouri} O.,    {Ma} C.-P.,  2009, \mnras, 398, 1858

\bibitem[\protect\citeauthoryear{{Metuki}, {Libeskind}, {Hoffman}, {Crain} \&
  {Theuns}}{{Metuki} et~al.}{2015}]{Metuki:2015}
{Metuki} O.,  {Libeskind} N.~I.,  {Hoffman} Y.,  {Crain} R.~A.,    {Theuns} T.,
   2015, \mnras, 446, 1458

\bibitem[\protect\citeauthoryear{{Naab} \& {Ostriker}}{{Naab} \&
  {Ostriker}}{2006}]{Naab:2006}
{Naab} T.,  {Ostriker} J.~P.,  2006, \mnras, 366, 899

\bibitem[\protect\citeauthoryear{{Nuza}, {Kitaura}, {He{\ss}}, {Libeskind} \&
  {M{\"u}ller}}{{Nuza} et~al.}{2014}]{Nuza:2014}
{Nuza} S.~E.,  {Kitaura} F.-S.,  {He{\ss}} S.,  {Libeskind} N.~I.,
  {M{\"u}ller} V.,  2014, \mnras, 445, 988

\bibitem[\protect\citeauthoryear{{Parry}, {Eke} \& {Frenk}}{{Parry}
  et~al.}{2009}]{Parry:2009}
{Parry} O.~H.,  {Eke} V.~R.,    {Frenk} C.~S.,  2009, \mnras, 396, 1972

\bibitem[\protect\citeauthoryear{{Pe{\~n}arrubia}, {G{\'o}mez}, {Besla},
  {Erkal} \& {Ma}}{{Pe{\~n}arrubia} et~al.}{2016}]{Penarrubia:2016}
{Pe{\~n}arrubia} J.,  {G{\'o}mez} F.~A.,  {Besla} G.,  {Erkal} D.,    {Ma}
  Y.-Z.,  2016, \mnras, 456, L54

\bibitem[\protect\citeauthoryear{{Penzo}, {Macci{\`o}}, {Baldi}, {Casarini},
  {O{\~n}orbe} \& {Dutton}}{{Penzo} et~al.}{2016}]{Penzo:2016}
{Penzo} C.,  {Macci{\`o}} A.~V.,  {Baldi} M.,  {Casarini} L.,  {O{\~n}orbe} J.,
     {Dutton} A.~A.,  2016, \mnras, 461, 2490

\bibitem[\protect\citeauthoryear{{Planck Collaboration}}{{Planck
  Collaboration}}{2014}]{Planck:2013}
{Planck Collaboration} 2014, \aap, 571, A16

\bibitem[\protect\citeauthoryear{{Rodriguez-Gomez} et~al.,}{{Rodriguez-Gomez}
  et~al.}{2017}]{Rodriguez-Gomez:2017}
{Rodriguez-Gomez} V.,  et~al., 2017, \mnras, 467, 3083

\bibitem[\protect\citeauthoryear{{Ruchti} et~al.,}{{Ruchti}
  et~al.}{2015}]{Ruchti:2015}
{Ruchti} G.~R.,  et~al., 2015, \mnras, 450, 2874

\bibitem[\protect\citeauthoryear{{Saglia}, {Fabricius}, {Bender}, {Montalto},
  {Lee}, {Riffeser}, {Seitz}, {Morganti}, {Gerhard} \& {Hopp}}{{Saglia}
  et~al.}{2010}]{Saglia:2010}
{Saglia} R.~P.,  {Fabricius} M.,  {Bender} R.,  {Montalto} M.,  {Lee} C.-H.,
  {Riffeser} A.,  {Seitz} S.,  {Morganti} L.,  {Gerhard} O.,    {Hopp} U.,
  2010, \aap, 509, A61

\bibitem[\protect\citeauthoryear{{Sales}, {Navarro}, {Theuns}, {Schaye},
  {White}, {Frenk}, {Crain} \& {Dalla Vecchia}}{{Sales}
  et~al.}{2012}]{Sales:2012}
{Sales} L.~V.,  {Navarro} J.~F.,  {Theuns} T.,  {Schaye} J.,  {White} S.~D.~M.,
   {Frenk} C.~S.,  {Crain} R.~A.,    {Dalla Vecchia} C.,  2012, \mnras, 423,
  1544

\bibitem[\protect\citeauthoryear{{Salomon}, {Ibata}, {Famaey}, {Martin} \&
  {Lewis}}{{Salomon} et~al.}{2016}]{Salomon:2016}
{Salomon} J.-B.,  {Ibata} R.~A.,  {Famaey} B.,  {Martin} N.~F.,    {Lewis}
  G.~F.,  2016, \mnras, 456, 4432

\bibitem[\protect\citeauthoryear{{Sawala}, {Frenk}, {Fattahi}, {Navarro},
  {Bower}, {Crain}, {Dalla Vecchia}, {Furlong}, {Helly}, {Jenkins}, {Oman},
  {Schaller}, {Schaye}, {Theuns}, {Trayford} \& {White}}{{Sawala}
  et~al.}{2014}]{Sawala:2014}
{Sawala} T.,  {Frenk} C.~S.,  {Fattahi} A.,  {Navarro} J.~F.,  {Bower} R.~G.,
  {Crain} R.~A.,  {Dalla Vecchia} C.,  {Furlong} M.,  {Helly} J.~C.,  {Jenkins}
  A.,  {Oman} K.~A.,  {Schaller} M.,  {Schaye} J.,  {Theuns} T.,  {Trayford}
  J.,    {White} S.~D.~M.,  2014, ArXiv 1412.2748

\bibitem[\protect\citeauthoryear{{Scannapieco}, {Creasey}, {Nuza}, {Yepes},
  {Gottl{\"o}ber} \& {Steinmetz}}{{Scannapieco}
  et~al.}{2015}]{Scannapieco:2015}
{Scannapieco} C.,  {Creasey} P.,  {Nuza} S.~E.,  {Yepes} G.,  {Gottl{\"o}ber}
  S.,    {Steinmetz} M.,  2015, \aap, 577, A3

\bibitem[\protect\citeauthoryear{{Scannapieco}, {White}, {Springel} \&
  {Tissera}}{{Scannapieco} et~al.}{2009}]{Scannapieco:2009}
{Scannapieco} C.,  {White} S.~D.~M.,  {Springel} V.,    {Tissera} P.~B.,  2009,
  \mnras, 396, 696

\bibitem[\protect\citeauthoryear{{Sohn}, {Anderson} \& {van der Marel}}{{Sohn}
  et~al.}{2012}]{Sohn:2012}
{Sohn} S.~T.,  {Anderson} J.,    {van der Marel} R.~P.,  2012, \apj, 753, 7

\bibitem[\protect\citeauthoryear{{Sorce}}{{Sorce}}{2015}]{Sorce:2015}
{Sorce} J.~G.,  2015, \mnras, 450, 2644

\bibitem[\protect\citeauthoryear{{Sorce}, {Courtois}, {Gottl{\"o}ber},
  {Hoffman} \& {Tully}}{{Sorce} et~al.}{2014}]{Sorce:2014}
{Sorce} J.~G.,  {Courtois} H.~M.,  {Gottl{\"o}ber} S.,  {Hoffman} Y.,
  {Tully} R.~B.,  2014, \mnras, 437, 3586

\bibitem[\protect\citeauthoryear{{Sorce}, {Gottl{\"o}ber}, {Yepes}, {Hoffman},
  {Courtois}, {Steinmetz}, {Tully}, {Pomar{\`e}de} \& {Carlesi}}{{Sorce}
  et~al.}{2016}]{Sorce:2016}
{Sorce} J.~G.,  {Gottl{\"o}ber} S.,  {Yepes} G.,  {Hoffman} Y.,  {Courtois}
  H.~M.,  {Steinmetz} M.,  {Tully} R.~B.,  {Pomar{\`e}de} D.,    {Carlesi} E.,
  2016, \mnras, 455, 2078

\bibitem[\protect\citeauthoryear{{Srisawat} \& {et al.}}{{Srisawat} \& {et
  al.}}{2013}]{Srisawat:2013}
{Srisawat} C.,  {et al.} 2013, \mnras, 436, 150

\bibitem[\protect\citeauthoryear{{Stewart}, {Bullock}, {Wechsler}, {Maller} \&
  {Zentner}}{{Stewart} et~al.}{2008}]{Stewart:2008}
{Stewart} K.~R.,  {Bullock} J.~S.,  {Wechsler} R.~H.,  {Maller} A.~H.,
  {Zentner} A.~R.,  2008, \apj, 683, 597

\bibitem[\protect\citeauthoryear{{Stinson}, {Bailin}, {Couchman}, {Wadsley},
  {Shen}, {Nickerson}, {Brook} \& {Quinn}}{{Stinson}
  et~al.}{2010}]{Stinson:2010}
{Stinson} G.~S.,  {Bailin} J.,  {Couchman} H.,  {Wadsley} J.,  {Shen} S.,
  {Nickerson} S.,  {Brook} C.,    {Quinn} T.,  2010, \mnras, 408, 812

\bibitem[\protect\citeauthoryear{{Tasitsiomi}, {Kravtsov}, {Gottl{\"o}ber} \&
  {Klypin}}{{Tasitsiomi} et~al.}{2004}]{Tasi:2004}
{Tasitsiomi} A.,  {Kravtsov} A.~V.,  {Gottl{\"o}ber} S.,    {Klypin} A.~A.,
  2004, \apj, 607, 125

\bibitem[\protect\citeauthoryear{{Tollerud}, {Boylan-Kolchin} \&
  {Bullock}}{{Tollerud} et~al.}{2014}]{Tollerud:2014}
{Tollerud} E.~J.,  {Boylan-Kolchin} M.,    {Bullock} J.~S.,  2014, \mnras, 440,
  3511

\bibitem[\protect\citeauthoryear{{Toth} \& {Ostriker}}{{Toth} \&
  {Ostriker}}{1992}]{Toth:1992}
{Toth} G.,  {Ostriker} J.~P.,  1992, \apj, 389, 5

\bibitem[\protect\citeauthoryear{{Tully}, {Courtois}, {Hoffman} \&
  {Pomar{\`e}de}}{{Tully} et~al.}{2014}]{Tully:2014}
{Tully} R.~B.,  {Courtois} H.,  {Hoffman} Y.,    {Pomar{\`e}de} D.,  2014,
  \nat, 513, 71

\bibitem[\protect\citeauthoryear{{van der Marel}, {Fardal}, {Sohn}, {Patel},
  {Besla}, {del Pino}, {Sahlmann} \& {Watkins}}{{van der Marel}
  et~al.}{2019}]{Marel:2019}
{van der Marel} R.~P.,  {Fardal} M.~A.,  {Sohn} S.~T.,  {Patel} E.,  {Besla}
  G.,  {del Pino} A.,  {Sahlmann} J.,    {Watkins} L.~L.,  2019, \apj, 872, 24

\bibitem[\protect\citeauthoryear{{van Dokkum} et~al.,}{{van Dokkum}
  et~al.}{2013}]{vanDokkum:2013}
{van Dokkum} P.~G.,  et~al., 2013, \apjl, 771, L35

\bibitem[\protect\citeauthoryear{{Wang}, {Mo}, {Yang}, {Zhang}, {Shi}, {Jing},
  {Liu}, {Li}, {Kang} \& {Gao}}{{Wang} et~al.}{2016}]{WangCS:2016}
{Wang} H.,  {Mo} H.~J.,  {Yang} X.,  {Zhang} Y.,  {Shi} J.,  {Jing} Y.~P.,
  {Liu} C.,  {Li} S.,  {Kang} X.,    {Gao} Y.,  2016, \apj, 831, 164

\bibitem[\protect\citeauthoryear{{Wang}, {Pearce}, {Knebe}, {Schneider},
  {Srisawat}, {Tweed}, {Jung}, {Han}, {Helly}, {Onions}, {Elahi}, {Thomas},
  {Behroozi}, {Yi}, {Rodriguez-Gomez}, {Mao}, {Jing} \& {Lin}}{{Wang}
  et~al.}{2016}]{Wang:2016}
{Wang} Y.,  {Pearce} F.~R.,  {Knebe} A.,  {Schneider} A.,  {Srisawat} C.,
  {Tweed} D.,  {Jung} I.,  {Han} J.,  {Helly} J.,  {Onions} J.,  {Elahi} P.~J.,
   {Thomas} P.~A.,  {Behroozi} P.,  {Yi} S.~K.,  {Rodriguez-Gomez} V.,  {Mao}
  Y.-Y.,  {Jing} Y.,    {Lin} W.,  2016, \mnras, 459, 1554

\bibitem[\protect\citeauthoryear{{White} \& {Rees}}{{White} \&
  {Rees}}{1978}]{White:1978}
{White} S.~D.~M.,  {Rees} M.~J.,  1978, \mnras, 183, 341

\bibitem[\protect\citeauthoryear{{Zavala}, {Avila-Reese}, {Firmani} \&
  {Boylan-Kolchin}}{{Zavala} et~al.}{2012}]{Zavala:2012}
{Zavala} J.,  {Avila-Reese} V.,  {Firmani} C.,    {Boylan-Kolchin} M.,  2012,
  \mnras, 427, 1503

\end{thebibliography}

\bsp

\appendix
\section{\texttt{METROC++} -- a parallel code for the computation of merger trees in cosmological simulations}
\label{app:a}
\texttt{METROC++} is an acronym that stands for \emph{MErger TRees On C++}. 
The name is reference to the infamous third subway line (Metro C) in Rome (where EC grew up)
and whose assembly (still undergoing, by the time this paper is being submitted) 
is taking an amount of time comparable to the formation of a dark matter halo.
It is a \texttt{C++11} code designed to compute merger trees in cosmological simulations, that uses 
\texttt{MPI-2.0 C} bindings for the implementation of distributed-memory parallelism. 
The code uses particle IDs to match haloes and is able to track them among non-consecutive snapshots in order to 
cope with the shortcomings of the halo finders when dealing with subhaloes. These are two of the three properties
suggested by \citet{Srisawat:2013}, the third one, namely the ability of smoothing over large fluctuations in halo mass,
has not been implemented in the code and has been done in the post-processing stage.
A practical description of the usage of the code can be found in the user's guide 
\footnote{https://github.com/EdoardoCarlesi/MetroCPP/blob/master/doc/UG.pdf}
that is provided with it. Here we will only deal with the main features of the algorithms and the physical assumptions of the code. \\

\subsection{Parallelization strategy}
The simulation volume is split into $N^3$ sub-cubes, where $N$ is a user-defined number that
should be chosen ensuring that $L / N > 2.0$\hMpc\ ($L$ is the box size).
This volume splitting is performed only in the case of full box cosmological simulations, as the \emph{zoom-in}
optimization of the code relies on a purely \emph{serial} version of the algorithm.
Each task is assigned a (comoving) sub-volume of the simulation box comprised of $\approx N /  N_{task}$ cells.
Halo properties with their particle ID content are then retrieved from two catalogs at the time
(corresponding to two redshifts $z_i$ and $z_{i+1}$, with $z_{i+1}>z_i$) and assigned to each task according to their position in space.
The cells at the edge of every sub-volume constitute a buffer zone that tasks communicate to their neighbours to ensure
that the haloes will be then consistently compared to \emph{all} of their potential progenitors.
While a larger grid cell (smaller $N$) might result in a better reconstruction of the trees, 
enlarging the list of potential progenitors, it also means that the buffers to be communicated among the tasks are larger, slowing
down the execution of the program and increasing the memory request.
On the other hand, smaller grid cells (larger $N$) lead to a speed-up of the code execution and reduce memory consumption at the 
expense of accuracy. It is up to the user to decide how to balance this trade-off.
In the present paper, we selected $N=48$ for the computation of the \std\
merger trees after checking for convergence, that is, ensuring that the results produced would not change for $N < 48$.\\

\subsection{Tree building and halo tracking}
After all haloes with their particle ID content have been correctly read and distributed, 
each task produces a \texttt{map}
\footnote{According to \texttt{http://www.cplusplus.com/} a map is  
\emph{an associative container that stores elements formed by a combination of a key value and a mapped value, 
following a specific order.} In practice, we can use a map as an array whose element are not to be called by 
integer values but rather by \emph{keys} of arbitrary type. Defining a \texttt{particleHaloMap} linking \texttt{particleID} 
to \texttt{haloID} we can quickly retrieve the ; e.g. if \texttt{particleID}$=12345$ and \texttt{haloID}$=6789$ then 
\texttt{particleHaloMap}$[12345]$ will return $6789$. This property makes the process of comparing very large 
arrays of particle IDs extremely quick, at the price of a larger memory requirements.} 
linking the particle ID to the \texttt{vector} of halo IDs to which it belongs to.
In this way, a single particle might be assigned to more haloes at the same time - e.g. a sub-halo and its host.
The maps are produced for both the $i$-th and $i+1$-th catalogs on each task and their particle ID content is compared to identify the 
halo connections, which are established only among objects sharing at least $N_{min}$ (user defined) particles.
For each halo $A$ at $z_i$ with $N_A$ particles, sharing
at least $N_{min}$ particles with $N^\prime$ haloes at $z_{i+1}$, 
we computing the merit function \citep[see][]{Knebe:2013}:

\begin{equation}
	\label{eq:m}
	\sum _{j=0} ^{N^\prime}
M(N_A, N_j) = \frac{N_{A,j}}{N_A, N_j}
\end{equation}

\noindent
where $N_j$ is the number of particles of the $j$-th possible progenitor.
The comparisons are carried for $z_i \rightarrow z_{i+1}$ (forwards) and $z_{i+1} \rightarrow z_i$ (backwards), 
and all the connections above $N_{min}$ are ranked using \Eq{eq:m} and sorted in descending order.
A halo $h_j$ is labeled as progenitor of $h_A$ if the latter maximizes the former's merit function, 
ensuring that haloes have a \emph{unique} descendant and they do not split.
If a halo is left with no likely progenitor, and provided it is composed by a number of particles above
a user-specified threshold, it is labeled as \emph{orphan} and 
the code still keeps track of its particle ID content for comparison during the following
snapshots. This enables to reconnect subhaloes with their
progenitor at a non-subsequent snapshot if the halo finder fails to identify it at any step, 
which typically happens when a subhalo is orbiting close to the center of mass of its host 
\citep{Avila:2014, Wang:2016}.
During the time steps in which the halo has been missing, the code replaces it with a token halo 
with exactly the same properties of the last valid descendant, until a proper descendant is can be identify.
After a (user specified) maximum number of steps, if no progenitor can be found for an orphan halo, the 
particle content is dropped and the merger tree is truncated.\\
At each step, once the pairwise catalog comparison has finished, the code runs a check on the 
halos in the buffer zone (which might have been assigned to a different descendant on each task) 
to ensure that they are correctly assigned to their a unique descendant. Then, the list of halos at $z_i$
are printed to a single output file together with their rank-ordered (by the merit function) descendants 
at $z_{i+1}$. \\
With the code it is also provided a series of scripts that allow to reconstruct the
merging histories reading from the \texttt{ASCII} output files 
and store them into a \texttt{SQL} database that can be queried using the 
$z=0$ ID of the halos.

\label{lastpage}

\end{document}